\documentclass[format=sigconf, screen=true]{acmart}

\AtBeginDocument{%
  }

\setcopyright{none}
\copyrightyear{2024}
\acmConference[40ème conférence sur la gestion des données (BDA)]{BDA}{October 21-24, 2024}{Orléans, France}

\usepackage{graphicx} 
\usepackage[pagewise]{lineno}
\usepackage{epsfig}
\usepackage{xcolor,colortbl}        
\usepackage{times}
\usepackage{fancyhdr,graphicx}
\usepackage{algorithm}
\usepackage{algpseudocode}
\usepackage{amsmath}
\usepackage{algorithm}
\usepackage{algpseudocode}
\usepackage{listings}
\usepackage{hyperref}
\usepackage{url} 
\usepackage{cleveref}

\newcommand{\ie}{\textit{i}.\textit{e}. }

\newcommand{\SimTr}{\textit{S}}
\begin{document}

\title{TISIS : Trajectory Indexing for SImilarity Search}
\author{Sara Jarrad, Hubert Naacke, Stéphane Gançarski}
\email{{sara.jarrad, hubert.naacke, stephane.gancarski}@lip6.fr}

\affiliation{%
  \institution{LIP6, Sorbonne Université}
  \city{Paris}
  \country{France}
}

\renewcommand{\shortauthors}{Sara Jarrad}
\begin{abstract}
Social media platforms enable users to share diverse types of information, including geolocation data that captures their movement patterns. Such geolocation data can be leveraged to reconstruct the trajectory of a user's visited Points of Interest (POIs). A key requirement in numerous applications is the ability to measure the similarity between such trajectories, as this facilitates the retrieval of trajectories that are similar to a given reference trajectory. 
 This is the main focus of our work.

Existing methods predominantly rely on applying a similarity function to each candidate trajectory to identify those that are sufficiently similar. However, this approach becomes computationally expensive when dealing with large-scale datasets. To mitigate this challenge, we propose \textit{TISIS}, an efficient method that uses trajectory indexing to quickly find similar trajectories that share common POIs in the same order.

Furthermore, to account for scenarios where POIs in trajectories may not exactly match but are contextually similar, we introduce \textit{TISIS*}, a variant of TISIS that incorporates POI embeddings. This extension allows for more comprehensive retrieval of similar trajectories by considering semantic similarities between POIs, beyond mere exact matches.

Extensive experimental evaluations demonstrate that the proposed approach significantly outperforms a baseline method based on the well-known Longest Common SubSequence (LCSS) algorithm, yielding substantial performance improvements across various real-world datasets.


\end{abstract}
\maketitle

\section{Introduction and motivation}
\label{sec:intro}
Social networks provide valuable information about user mobility and behavior, including geolocation data. 

This information can be used to construct a user's trajectory of visited Points of Interest (POIs).
A trajectory is a sequence of time-stamped POIs that describes the user's movements in chronological order.
A crucial requirement for many applications is measuring the similarity between these trajectories, which can help \textit{retrieve trajectories similar to a given query $q$, where $q$ is a sequence of POIs)}. This is the main focus of our work.

Trajectory retrieval is important in a number of applications, such as analyzing the travel patterns of users as a function of the POIs they visit. This can help to improve traffic management. 
Other examples include classification of similar users and POI/trajectory recommendation.

Many research studies have focused on quantifying the similarity between two trajectories using established methods such as Longest Common SubSequence (LCSS) \cite{lcss_classique1,lcss2,lcss_classique3}, Edit Distance on Real Sequence (EDR)\cite{EDR}, and Edit Distance with Real Penalty (ERP)\cite{ERP}.
However, a major limitation of these approaches is their computational cost, especially when applied to large datasets. Dynamic Time Warping (DTW)\cite{DTW1,DTW2} is an efficient method for finding similarities in time series, but cannot be applied to our sequences because we don't consider distances between POIs.
LCSS computes the length of the longest common subsequence of points within two sequences, where two compared POIs are considered to match only if they are identical.

A variant of the LCSS algorithm, denoted as $LCSS_\epsilon^{geo}$ \cite{Discovering_Similar_Multidimensional_Trajectories,Indexing_multi_dimensional_time_series,Review_on_trajectory_similarity_measures}, extends the original approach by considering POIs that are represented as points with coordinates in a two-dimensional (x, y) plane.

In this method, a match between two points is established when the difference in both the x and y coordinates is smaller than a predefined threshold, $\epsilon$. However, this form of spatial relaxation introduces significant limitations. It indiscriminately matches POIs solely based on spatial proximity, which can lead to erroneous results. Specifically, two POIs that are geographically close may still be contextually unrelated, making such spatial matching unreliable.

In our study, we utilize the computed length of the LCSS to assess the similarity between two sequences, addressing the contextual relevance of the matched POIs.

Although LCSS-based similarity is not a metric (\ie it does not satisfy the triangular inequality), it is widely regarded as one of the most effective measures due to its robustness against various transformations that may be applied to trajectories, such as re-sampling (adding points to trajectories) and handling data noise (introducing outliers in the trajectories)~\cite{Review_on_trajectory_similarity_measures,survey_of_trajectory_distance_measures,An_effectiveness_study_on_trajectory_similarity_measures}. 

Despite these advantages, LCSS can be computationally expensive when applied to large datasets of sequences. In a naive approach (baseline) for identifying trajectories similar to a given query trajectory $q$, the LCSS between $q$ and each candidate trajectory is calculated exhaustively. This inefficiency in processing time represents a significant challenge, and is one of the primary issues we address in this work.

A second issue we tackle is that exact matching between POIs can be overly restrictive, potentially resulting in very few or no matching trajectories. In many practical scenarios, users may not require exact POI matches but instead seek trajectories whose POIs are contextually similar to those in the query.
To address these challenges, we propose \textit{TISIS} (Trajectory Indexing for SImilarity Search). TISIS is an efficient method that employs a trajectory indexing approach to accelerate the search for similar trajectories. This approach achieves the same results as the LCSS-based baseline method (see~\Cref{sec:LCSS-based search}), while significantly improving search speed. Moreover, TISIS allows users to specify a desired similarity threshold, defined as the number of POIs in common with the query, in the same sequential order.

To further enhance the retrieval of similar trajectories, even when their POIs do not exactly match but are contextually similar, we introduce \textit{TISIS*}. TISIS* is a variant of TISIS that leverages POI embeddings generated by the Word2Vec (W2V) language model~\cite{efficient_word_estimation_using_w2v}, enabling a broader set of results compared to the exact matching performed by the original TISIS method.

In the rest of this paper, we introduce a background on LCSS and related work in ~\Cref{sec:Background} and ~\Cref{sec:related-work}  . Then we present the exact trajectory indexing in~\Cref{sec:exact-search}. After that, we introduce the trajectory indexing based on POIs embeddings in~\Cref{sec:embeddings-index-search}. Experimental validation and results are presented in~\Cref{sec:experiences}. Finally we conclude the paper, and propose future work in~\Cref{sec:conclusion}.

\section{Background}
\label{sec:Background}
The algorithm described in~\Cref{alg:LCSS} has been detailed in \cite{lcss_classique1}, to compute the size of the longest common subsequence between two sequences. 
This size is later used to compute the similarity between trajectories.
We extend the original version by adding the POI matching function as a parameter.

\begin{algorithm}[htbp]
\caption{LCSS size}
\label{alg:LCSS}
\begin{algorithmic}[1]
\Require{$q$: query trajectory, $t$: candidate trajectory, $equals$: POI matching function} 
    \Statex
\Function{LCSS}{$q, t, equals$}
    \State $ \forall i\in [1, |q|], j\in [1, |t|], ~ M[i][j] \gets 0$ ~ // Initialize the similarity matrix

    \For{$i$ in $[1, |q|]$, $j$ in $[1, |t|]$}
            \If{$equals(q[i - 1],  t[j - 1])$}
                \State $M[i][j] \gets M[i - 1][j - 1] + 1$
            \Else
                \State $M[i][j] \gets \max(M[i - 1][j], M[i][j - 1])$ 
         \EndIf
    \EndFor
    \State \textbf{return} $M[|q|][|t|]$  \quad // length of the LCSS 
\EndFunction
\end{algorithmic}
\end{algorithm}

$LCSS(q,t)$ denotes $LCSS(q, t, = )$ \ie  the default $equals$ is the equality function and two POIs being compared only match if they are the same.

\begin{example}
\label{ex:lcss}
Consider query $q = [A, \textbf{D}, B, \textbf{E}, \textbf{C}]$, and a trajectory $t=[F, \textbf{D}, G, \textbf{E}, H, \textbf{C}, A]$, then the longest common sub-sequence is $[\textbf{D}, \textbf{E}, \textbf{C}]$ and $LCSS(q,t)$ = 3.
\end{example}

\subsection{Problem Statement}
\label{sec:problem}
Let  $T=(t_1, \cdots, t_n)$ be a set of trajectories. 
Let $q,t \in T$ and  $LCSS(q,t)$ be the size of the LCSS between $q$ and $t$ defined in~\Cref{alg:LCSS}.

We consider a trajectory $t$ to be similar to $q$ for a given similarity threshold $\SimTr\in [0,1]$ if its LCSS contains at least $p=\lceil |q| \times \SimTr \rceil$ POIs. In other words, the ratio of the sizes of $q$ and LCSS(q,t) must be greater than $\SimTr$. We denote $q \approx_{\SimTr} t$ such a similarity property defined by:
$$ q \approx_{\SimTr} t \equiv \frac{ LCSS(q , t) }{|q|} \ge \SimTr$$

We consider a user submitting a query composed  of an input trajectory $q$, and  
a similarity threshold $\SimTr$.
We aim to retrieve the set $T'(q)$ of all trajectories $T$ that are similar enough to $q$. 
We have $T'(q) = \{ t\in T  | q \approx_{\SimTr} t \}$.

\begin{example}
\label{ex:sim}
Let us consider a query trajectory $q = [A, B, C, D, E]$ and two candidate trajectories: $t_1 = [K, \textbf{A}, F, \textbf{D}]$ and $t_2$ = [ $M$, $O$, \textbf{A}, \textbf{B}, $F$, \textbf{C}, P, \textbf{E} ]. Suppose the user specifies a similarity threshold $S = 0.6$, corresponding to $p = 3$. 
As per the definitions provided above, trajectory $t_2$ is considered similar to $q$ under the given threshold $S$, as $\frac{\text{LCSS}(q, t_2)}{|q|} = \frac{4}{5} \geq S$. In contrast, trajectory $t_1$ does not meet the similarity threshold.
\end{example}

\subsection{LCSS-based baseline }
\label{sec:LCSS-based search}
As mentioned before, to determine the trajectories that are similar to a given query $q$, the LCSS between $q$ and all other trajectories is calculated (\Cref{alg:LCSS-based search}). Thus, the LCSS function defined in~\Cref{alg:LCSS} is invoked $|T|$ times.

\begin{algorithm}[htbp]
\caption{LCSS-based search}
\label{alg:LCSS-based search}
\begin{algorithmic}[1]
\Require{$q$: query trajectory, $S$: threshold} 
    \Statex
\Function{LCSS\_search}{$q, \SimTr$}
    \State{$result \gets \emptyset $, \quad $p\gets \lceil |q| \times \SimTr \rceil$} 
    \For{$t$ in $T$}
            \If{$LCSS(q, t, equals) \ge p$}
                \State $result \gets t$
         \EndIf
    \EndFor
    \State \textbf{return} $result$  
\EndFunction
\end{algorithmic}
\end{algorithm}
On lines 4 and 5 of~\Cref{alg:LCSS-based search} we select a trajectory if it contains at least $p$ POIs.
This process takes a long time, and we are looking to reduce it. \Cref{alg:LCSS-based search} serves as a baseline with which we will compare ourselves throughout the paper. 
Our aim is to propose a new approach to obtain similar trajectories faster than this baseline.
The notations frequently used throughout this paper are summarised in~\Cref{tab:symbols}.

\begin{table}[]
\centering
\caption{Notations used for similarity search}
\begin{tabular}{ p{0.15\linewidth}  p{0.75\linewidth} }
         \textbf{Notation} & \textbf{Description} \\  
         \hline
        $q$ & A query trajectory  \\ 
         $POI$ &  Point Of Interest \\ 
         $T$ & Set of trajectories \\
         $T'(q)$ & Set of all trajectories that are similar to the query $q$\\
         $q \approx_{\SimTr} t$ &  a trajectory $t$ is similar to $q$ for a given threshold $S$\\
         $pos(p_i,t)$ & The position of POI $p_i$ in a trajectory $t$\\ 
         $C_{|q|,p}$ & the set of $C_p^{|q|}$ combinations of size $p$ in the query $q$ \\
         $\overline{p_i}$ &  An index that associates each element $p_i$ with the trajectories that contain it\\
         $\overline{p_ip_j}$ &  An index that associates each pair of elements $p_ip_j$ with the trajectories that contain it \\
         $ sim_\epsilon(p_i, p_j)$ &  $p_i$ and $p_j$ are $\epsilon$-similar\\
          $\bar{\bar{p_i}}$ & An index that associates each POI $p_i$ with the set of trajectories that pass through a point $p_j$ $\epsilon$-similar to $p_i$ \\
        
        \hline
        \end{tabular}
\label{tab:symbols}
\end{table}

\section{Related work}
\label{sec:related-work}
There are  several studies that address the problem of finding similar trajectories. For instance \cite{Review_on_trajectory_similarity_measures}  presents  different categories of similarity measures, with the advantages and disadvantages of each in terms of noise, time shift (when an element of one trajectory is shifted in time to match an element of another trajectory) and trajectory length. These measures are classified according to whether they are spatial {\em i.e.} ignoring the temporal aspect, or spatio-temporal, taking into account both the spatial and temporal dimensions of motion. Spatial measures include Euclidean, Hausdroff, and Fréchet distances.
The first does not allow to compare trajectories of different sizes and is not robust to noise. The second and third are based solely on the geographical shape of the trajectories being compared, and are not robust to data noise or time shifts.
There are other similarity measures, such as DTW, EDR and ERP, which are among the most well-known, but unfortunately they are also not robust to noise. 

In  \cite{Review_on_trajectory_similarity_measures}, LCSS is considered one of the best measures because it is robust to almost all transformations compared to the other measures. 

\cite{survey_of_trajectory_distance_measures,An_effectiveness_study_on_trajectory_similarity_measures} compare the same measures, but categorizes them differently depending on whether they are spatio-temporal or sequence-only measures, and whether they are discrete or continuous. 
To compare the effectiveness of each of these measures, they are tested on the data by applying transformations such as adding/deleting sample points, adding noise (outliers), and shifting trajectories. They then analysed which measures dealt well with these problems. 
They found that DTW, LCSS, and ERP were the best measures for dealing with most of the transformations proposed in the paper.

A comparative study has also been proposed by \cite{Trajectory_similarity_measures}, which examines the four similarity measures LCSS, Fréchet, DTW and EDR. This paper highlights the differences between these four measures using real data and also shows some popular applications of these measures.
They find a strong positive correlation between Fréchet and DTW, and a strong negative correlation between LCSS and EDR.

Apart from the mobility field, LCSS is used in biology for DNA alignment. Other methods which have the same purpose as LCSS are, local alignment \cite{alignement_local1,alignement_local2}, global alignment \cite{alignement_global1,alignement_global2}, multiple alignment \cite{alignement_multiple}, and BLAST \cite{Blast}.
Local alignment identifies common elements between  two sequences, even if they do not overlap over their entire length. As for the global alignment, the objective is to maximize the overall similarity through the sequence alignment by taking into consideration the gaps between elements, while preserving the order of these elements. Multiple alignment aligns multiple sequences simultaneously to find common motifs among them all. Finally, Blast is used to find regions of similarity between biological sequences based on local alignment.
All of these measures are effective, but they are all complex on a large scale.

In summary, there are a variety of methods, using different approaches to solve trajectory similarity search problem, and the works cited explore the differences between them. All methods have advantages and disadvantages. However, according to the experimental validation used in \cite{Review_on_trajectory_similarity_measures,survey_of_trajectory_distance_measures,An_effectiveness_study_on_trajectory_similarity_measures},
LCSS remains one of the most robust to various mentioned transformations. Thus we base our work on this measure. 

The issue of those cited works is that they do not address the cost problem. In fact, these methods are effective when applied to a small dataset. Once scaled up, they become very costly and time consuming.

\cite{LCIS} use efficient algorithms for finding a longest common increasing subsequence among $m$ sequences. 
However, they assume that the sequences of points they use are increasing, \ie the points are ordered by increasing values. In our case, if the points were ordered by their values, the POI visit order would be lost. 
Therefore, the solution is not applicable to our problem.

\cite{LCS_timeSeries} use a dual-match method for matching sub-sequences. Their aim is to find all the subsequences (from a very large data sequence) that are similar to a given input query. The authors of \cite{LCS_timeSeries} present a complex solution, by slicing the data sequence and the query into windows 
then representing it in the form of multi-dimensional vectors, then applying the LCSS using a parameter to match intervals in the time dimension. 
They apply their method to time series data:
their index is based on the data value and the elapsed time between two consecutive points, which is not the case for us.

Some other recent works such as~\cite{approximation_LCSS1,approximation_LCSS2} have proposed solutions to approximate LCSS in linear time (instead of quadratic time as for~\Cref{alg:LCSS}) with respect to the size of sequences.
This makes the computation time of LCSS faster. However, the problem with these approaches is that they do not give the exact results provided by LCSS, but rather an approximation using probability to make a tradeoff between approximation accuracy and running time.
Whereas these works compute LCSS whatever the size of the longest common sub-sequence,
we study a different problem where the minimal size of the sub-sequence to find is specified by the user.

\section{Exact trajectory indexing}
\label{sec:exact-search}
This section introduces the main contribution of this work: an indexing method designed for the efficient retrieval of similar trajectories.

Let $T = (t_1, \dots, t_n)$ represent a set of trajectories, and let $T'(t)$ denote the subset of trajectories in $T$ that are sufficiently similar to a given trajectory $t$, as formally defined in \Cref{sec:problem}. The proposed method employs trajectory indexing to efficiently identify $T'(t)$. 

The key advantage of this indexing approach, compared to the baseline method outlined in \Cref{alg:LCSS-based search}, which performs pairwise comparisons of each trajectory in $T$ with $t$, is its ability to reduce the number of candidate trajectories. By narrowing down the search space, the indexing method is expected to significantly improve retrieval speed.

\subsection{Definition of trajectory index}
\begin{definition}[Single POI index]
\label{defs:index_by_single_poi} 
The trajectory index $\textit{1P}: p_i \mapsto \bar{p_i}$ associates each point of interest (POI) $p_i$ with the set $\bar{p_i}$, which consists of all trajectories that pass through $p_i$. This set is formally defined as:
$$
\bar{p_i} = \{ t \in T \mid p_i \in t \}
$$
\end{definition}

\begin{definition}[POI pair index]
\label{def:index_2P}
The trajectory index  $\textit{2P}: (p_i, p_j) \mapsto \overline{p_ip_j}$ 
associates a pair of POIs $(p_i, p_j)$ with the set $\overline{p_ip_j}$ of trajectories that pass through $p_i$ then through $p_j$ (\ie $p_i$ precedes $p_j$)  and defined by
$$\overline{p_ip_j} = \{ t \in T | p_i,p_j \in t ~ \textrm{and} ~ pos(p_i, t) < pos(p_j, t) \}$$ where $pos(p, t)$ is the position of POI $p$ in trajectory $t$.
\end{definition}

The index structure is implemented as a dictionary, where each point of interest (POI) or pair of POIs is mapped to the set of corresponding trajectories.

\subsection{Index-based similarity search.}
\label{sec:index_based_similarity_search}
We first retrieve the $\bar{p_i}$ trajectories for each POI in the query $q$.
Since the $\SimTr$ threshold indicates that the user expects to obtain trajectories that share at least $p$ POIs with $q$, we generate all the sub-trajectories of $q$ of size $p$.
For each sub-trajectory, we obtain candidate results by computing the intersection of the sets $\bar{p_i}$ corresponding to the POIs in that sub-trajectory.
Finally, we verify whether the POIs in the sub-trajectory appear in the right order in the candidate trajectory, otherwise the candidate is filtered out.

We provide a detailed explanation of~\Cref{alg:sim_traj} which retrieves similar trajectories.
The algorithm takes as input a query $q$ and a similarity threshold $\SimTr$.

\begin{algorithm}[H] 
\caption{Find similar trajectories}
\label{alg:sim_traj}
\begin{algorithmic}[1]
    \Require{$q$: query trajectory, $S$: threshold} 
    \Statex
    \Function{Similar\_trajectories}{$q, \SimTr$} 
        \State{$result \gets \emptyset $, \quad $p\gets \lceil |q| \times \SimTr \rceil$}
        \State{$C_{|q|,p}\gets $ the set of $C_p^{|q|}$ combinations of size p in q}
        \For{$combi \in C_{|q|,p}$ }
            \State{$candidates \gets \bigcap\limits_{i \in combi} \bar{p_i} $}
            \For{$c \in candidates$ } 
                \If{$c \notin result$ and $same\_order(c, combi)$}
                     \State{$result \gets result \cup \{ c\} $ }
                 \EndIf
            \EndFor
        \EndFor
  
        \State \Return {$result$}
    \EndFunction
\end{algorithmic}
\end{algorithm}

On line 2, the minimum number of POIs, $p$, is computed from the user's threshold $\SimTr$.
On line 3, we generate the set of $C_p^{|q|}$ combinations of size $p$ from the query trajectory $q$.
For each combination, denoted as $combi$, we compute the set of candidate trajectories that contain all the POIs in $combi$ by intersecting the sets $\bar{p_i}$ for which $i \in combi$.  
Finally on lines 7-9, we verify that the order of the POIs in each candidate trajectory matches the order in $combi$ (as detailed in~\Cref{alg:filtering}). 

Algorithm~\ref{alg:filtering} takes as input a candidate trajectory $c$, a combination $combi$, and a POI matching function (with POI equality by default).
The algorithm iterates over the POIs in $c$ and $combi$, comparing them consecutively:
If the POIs at the current positions $i$ in $c$ and $j$ in $combi$ are equal, the match count is incremented, and the algorithm proceeds to compare the POIs at positions $i+1$ and $j+1$. Otherwise, \ie if the POIs do not match, the next POI of the candidate trajectory (at position $i+1$) is compared with the current POI of $combi$ (at position $j$).
After going through all the POIs in $combi$, the algorithm returns $True$ if all POIs of $combi$ were matched, \ie if the number of matches $m$ equals the size of $combi$.

The source code for the implementation of this solution is available at \footnote{\url{https://gitlab.lip6.fr/jarrad/tisis/-/tree/main}}.

\begin{algorithm}[H]
\caption{Check identical POIs order}
\label{alg:filtering}
\begin{algorithmic}[1]
\Require{$c$: candidate trajectory, $combi$: sub-trajectory, $equals$: POI matching function } 
\Statex
\Function{same\_order}{$c, combi, equals$} 
    \State{$i \gets 0$, \quad $j \gets 0$, \quad $m \gets 0$}
    \While{$i < |c| $ \textbf{and} $j < |combi|$}
        \If{$equals(c[i], combi[j])$}
            \State{increment $j$,  ~ increment $m$}
        \EndIf
        \State increment $i$
    \EndWhile
    \State \textbf{return} $(m = |combi|)$
\EndFunction
\end{algorithmic}
\end{algorithm}

\subsection{Indexing trajectories by pairs of POIs}
\label{sec:index_by_paire_pois}
To enhance the performance (\ie reduce the response time) of similar trajectory retrieval, we extend the above indexing method to access trajectories containing not just a single (POI), but pairs of POIs. Since a POI pair index (2P index, see~\Cref{def:index_2P}) is more selective than a single POI index (1P index), we expect to get a performance benefit at the price of a larger index that is longer to build. 
However, the index creation time is considered to be amortized along the index usage period, thus this solution is considered  beneficial as long as it improves trajectory retrieval time.
Assuming a 2P index, we can access the set $\overline{p_ip_j}$ of trajectories that pass through $p_i$ followed by $p_j$ (\ie $p_i$ precedes $p_j$).
We adapt~\Cref{alg:sim_traj} to access trajectories through the POI pair index. Specifically, on line 5, the index is probed for every pair of consecutive POIs in $combi$.  Line 5 is modified as follows:

$$candidates \gets \bigcap\limits_{i,j \in combi, ~ pos(j, combi) = pos(i, combi) +1} \overline{p_ip_j} $$

\section{Trajectory indexing based on POI embeddings}
\label{sec:embeddings-index-search}
As discussed in~\Cref{sec:intro}, both the LCSS-based baseline and TISIS are effective in finding similar trajectories, but they rely on exact point matching. In some cases, this constraint can be too restrictive, leading to a small or even empty result set, particularly for large queries.

To relax this constraint, we propose a method based on POI embeddings generated using W2V model \cite{efficient_word_estimation_using_w2v}. The principle of W2V is that if two words frequently co-occur in input sentences, the model infers that they share a similar context and represents them using close numerical vectors. 
In our scenario, we can think of the words as being the POIs and the sentences as being the sequences containing these POIs. We assume that trajectories are similar not only when their POIs match, but also when the contexts of these POIs are close in the embedding space.

Based on this approach, we introduce TISIS*, an extension of TISISthat leverages embeddings. This variant takes into account a relaxation threshold specified by the user, and compares trajectories by exploiting contextual similarity.
This method provides more results than TISIS. Those additional trajectories do not necessarily pass through the query points but through points that are close to them in the vector space of the embeddings, {\em i.e.} they are {\em contextually} close.
specified

\subsection{Contextual similarity  and trajectory index}
\label{sec:contextual_sim}

Let $p_i$ and $p_j$ two POIs, and $p'_i$ and $p'_j$ their respective embedding obtained by Word2Vec (W2V)~\cite{efficient_word_estimation_using_w2v}.

\begin{definition}[POI $\epsilon$-similarity]
\label{defs:POI_sim}
We say that $p_i$ and $p_j$ are $\epsilon$-similar if the cosine between $p'_i$ and $p'_j$ is above a similarity threshold $\epsilon$ given by the user. We define $sim_\epsilon$ as
$$ sim_\epsilon(p_i, p_j) \equiv cosine(p'_i,p'_j)>= \epsilon $$
Obviously, if $p_i$ and $p_j$ are the same point, their cosine is equal to 1 and thus they are similar.
\end{definition}

The similarity between two trajectories is defined in~\Cref{sec:problem}, but considering a contextual version of LCSS, denoted $LCSS(q,t, sim_\epsilon)$, where two points match if they are $\epsilon$-similar, as proposed in~\Cref{sec:intro}.

We now generalize the definition of the index.

\begin{definition}[Contextual trajectory index]
\label{defs:contextual_index}
The contextual trajectory index  $\textit{CTI}: p_i \mapsto \bar{\bar{p_i}}$ 
associates each POI $p_i$ with the set $\bar{\bar{p_i}}$ of trajectories that pass through a point $p_j$ $\epsilon$-similar to $p_i$ and defined by 
$$\bar{\bar{p_i}} = \{ t \in T | \exists p_j \in t,  ~s.t. ~ sim_\epsilon(p_i, p_j) \}$$
\end{definition}

\subsection{Trajectory search based on POI embeddings}
This section details the method we designed to integrate POI embeddings in trajectory search.

Word2Vec (W2V)~\cite{efficient_word_estimation_using_w2v} is a language model designed to capture the semantic relationships in human language, which consists of words and sentences. W2V extracts information from sentences by representing each word as a dense numerical vector within a high-dimensional space. These vectors, known as \textit{embeddings}, are learned through a feed-forward neural network. By associating a dense vector with each word, W2V is able to capture the contextual meaning of words, such that words with similar contexts are represented by vectors that are close to one another in the embedding space.

In this study, we treat POIs as words and trajectories, \ie sequences of POIs, as sentences. By providing all sentences (trajectories) as input to the W2V model, we obtain vector representations for each word, referred to as POI embeddings, as the output.

These embeddings are used to create the contextual trajectory index $\bar{\bar{p_i}}$ and to identify similar trajectories as outlined in~\Cref{alg:sim_traj}, provided that line 5 becomes $$candidates \gets \bigcap\limits_{i \in combi} \bar{\bar{p_i}} $$
 
The step to create combinations of $p$ points (line 3 of~\Cref{alg:sim_traj}) remains the same. 
On line 7, the POI order check takes into account $\epsilon$-similarity POI matching by invoking $same\_order$ with $sim_\epsilon$ as the POI matching function.

It is important to note that the size of $\bar{\bar{p_i}}$ set introduced in~\Cref{defs:contextual_index} depends on the selected $\epsilon$ similarity threshold. 
The lower the threshold, the more similar trajectories will be found.
Our experiments will investigate the impact of $\epsilon$ on the result size, specifically the number of additional trajectories found.

\section{Experimental validation}
\label{sec:experiences}
In this section, we present the experimental evaluation of our proposed solution and compare its efficiency with the baseline. We begin by describing the datasets used and outlining the experimental methodology for both exact trajectory indexing and indexing based on POI embeddings. Subsequently, we analyze and discuss the results obtained.
All experiments were conducted on a Dell PowerEdge R440 server running Linux, equipped with 370 GB of RAM, and the implementation was carried out in Python. The index structure was implemented using a dictionary that maps each POI (or pair of POIs) to a set of trajectories 

For the embedding-based solution, we additionally rely on the multidimensional index provided by the W2V model, which retrieves POIs similar to a given POI in a 10-dimensional vector space.

\subsection{Data preparation}
\label{sec:dataset}
We conduct our experiments on three datasets: Foursquare\cite{foursquare_dataset} and Gowalla\cite{gowalla} on New York city, and YFCC\cite{yfcc} on France. The aim is to demonstrate the effectiveness of our method across different datasets.

Foursquare and Gowalla come with predefined POIs, but YFCC does not, so we use the method described in \cite{Embedding_Enhanced_Similarity_Metrics_for_Next_POI_Recommendation} to define the POIs.

In this study, we consider only POIs that have been visited more than 15 times. POIs with low visit frequencies are excluded as they do not contribute effectively to our study; trajectories passing through such POIs are sparse, resulting in insufficient similar trajectories. Our goal is to construct trajectories from fairly visited POIs to find a reasonable number of neighbours for each trajectory containing a common subsequence of POIs.

\begin{figure}
    \centering
    \includegraphics[width=\linewidth]{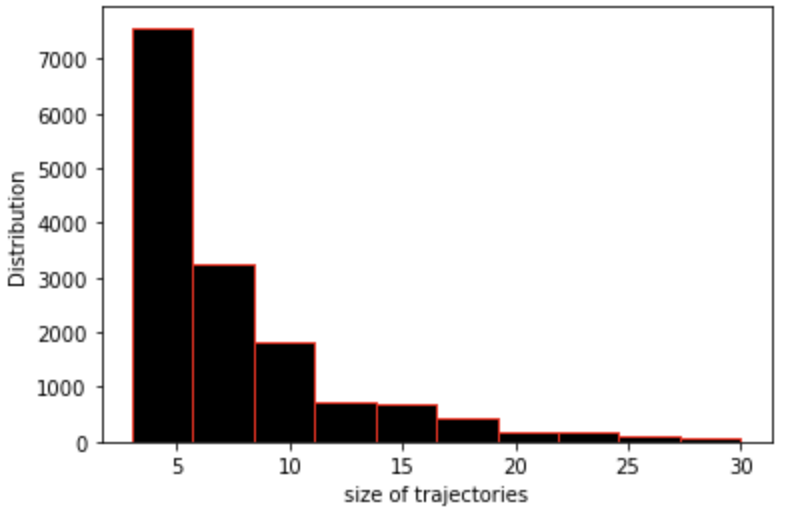}
    \caption{Foursquare}
    \label{fig:distribution_foursquare}
\end{figure}

\begin{figure}
    \centering
    \includegraphics[width=\linewidth]{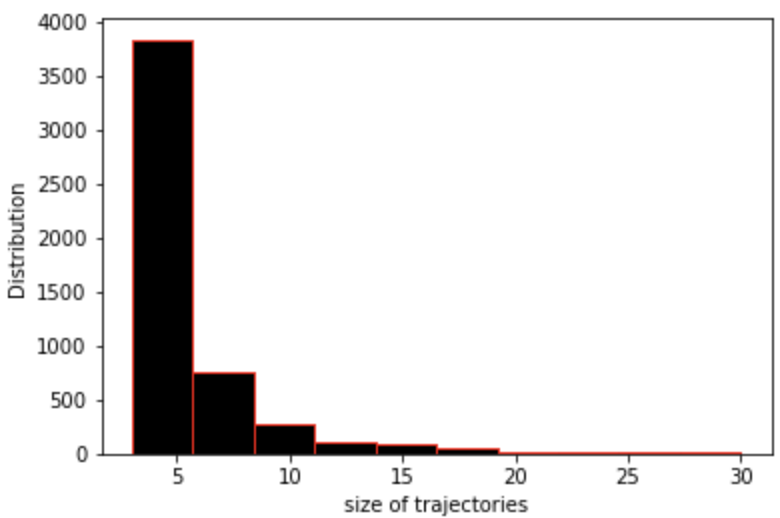}
    \caption{Gowalla}
    \label{fig:distribution_gowalla}
\end{figure}

\begin{figure}
    \centering
    \includegraphics[width=\linewidth]{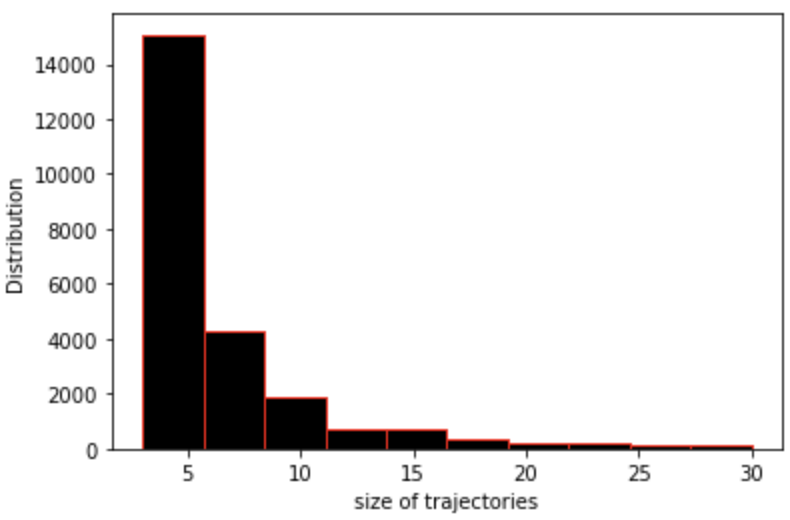}
    \caption{YFCC}
    \label{fig:distribution_yfcc}
\end{figure}

After selecting the POIs, we construct user trajectories on a daily basis (from 0:00 am to 11:59 pm).
Then, we keep only trajectories with a realistic size, {\em i.e.} in [3,30], 10,087 trajectories using Foursquare, 23,698 trajectories using YFCC, and 5,186 trajectories using Gowalla.
We analyse the distribution of trajectories sizes for the three datasets in~\Cref{fig:distribution_foursquare}, \ref{fig:distribution_gowalla}, and \ref{fig:distribution_yfcc}. The average trajectory size is 5 for Foursquare, 6 for Gowalla, and 5 for YFCC.
In the context of mobility data, trajectories tend to be relatively short. However, our approach can be generalized and applied to other domains where sequences are, on average, longer.

\subsection{Exact trajectory indexing: experimental methodology}
\label{sec:experiences_recherche_exact}
For our experiments, we use the three datasets described in section~\ref{sec:dataset}, and compare our proposed solution with the LCSS-based baseline. 
We use the trajectories in the dataset as queries. 
For each query, we retrieve all the similar trajectories, with various similarity thresholds ranging from 0.1 to 1, as well as the time taken to compute them.
We focus on the effect of query size on response time, and report the average response times for each query size.

As outlined in~\Cref{sec:index_based_similarity_search}, TISIS relies on the calculation of $C_p^{|q|}$ combinations of $p$ POIs from  $q$.
As the number of combinations gets larger, TISIS performs slower. 
Therefore, we investigate the worst-case scenario where $C_p^{|q|}$ is maximized, \ie $p = |q|/2$. Accordingly, we set the threshold \SimTr~to 0.5 in all the following experiments.

The experimental methodology described here applies to indexing trajectories by a single POI (described in~\Cref{sec:index_based_similarity_search}), and also to indexing trajectories by pairs of POIs (described in~\Cref{sec:index_by_paire_pois}). We carry out this experiment on Foursquare dataset in the~\Cref{sec:Time as a function of query size}, and subsequently compare the results with those obtained with the Gowalla and YFCC datasets in~\Cref{sec:comparaison_datasets}.
\subsection{Trajectory indexing based on POI embeddings: Experimental methodology}
\label{sec:experiences_embeddings}
In this section, we perform experiments using the Foursquare dataset and compare the results with those obtained using TISIS (exact method) on the same dataset. The aim is to demonstrate that TISIS* which is based on embeddings produces more interesting results than TISIS, thanks to the contextual similarity. We trained W2V with the following parameters: vector\_size=10, epochs = 5, window=5, and left all other parameters at their default values.
The user threshold is still $\SimTr = 0.5$ and  $\epsilon$ is in  $[0.65, 0.7, 0.75, 0.8, 0.85, 0.9, 0.95, 1]$.

\subsection{Results}
\label{sec:results and discussion}
\subsubsection{Exact indexing with 1 POI results}
\label{sec:results}

\begin{figure}
    \centering
    \includegraphics[width=\linewidth]{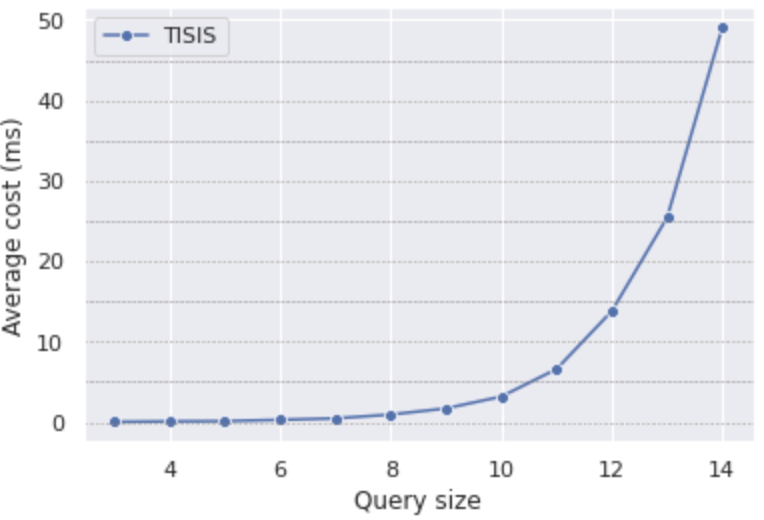}
    \caption{TISIS response time for small query sizes}
    \label{fig:tisis_small_sizes}
\end{figure}

\begin{figure}
    \centering
    \includegraphics[width=\linewidth]{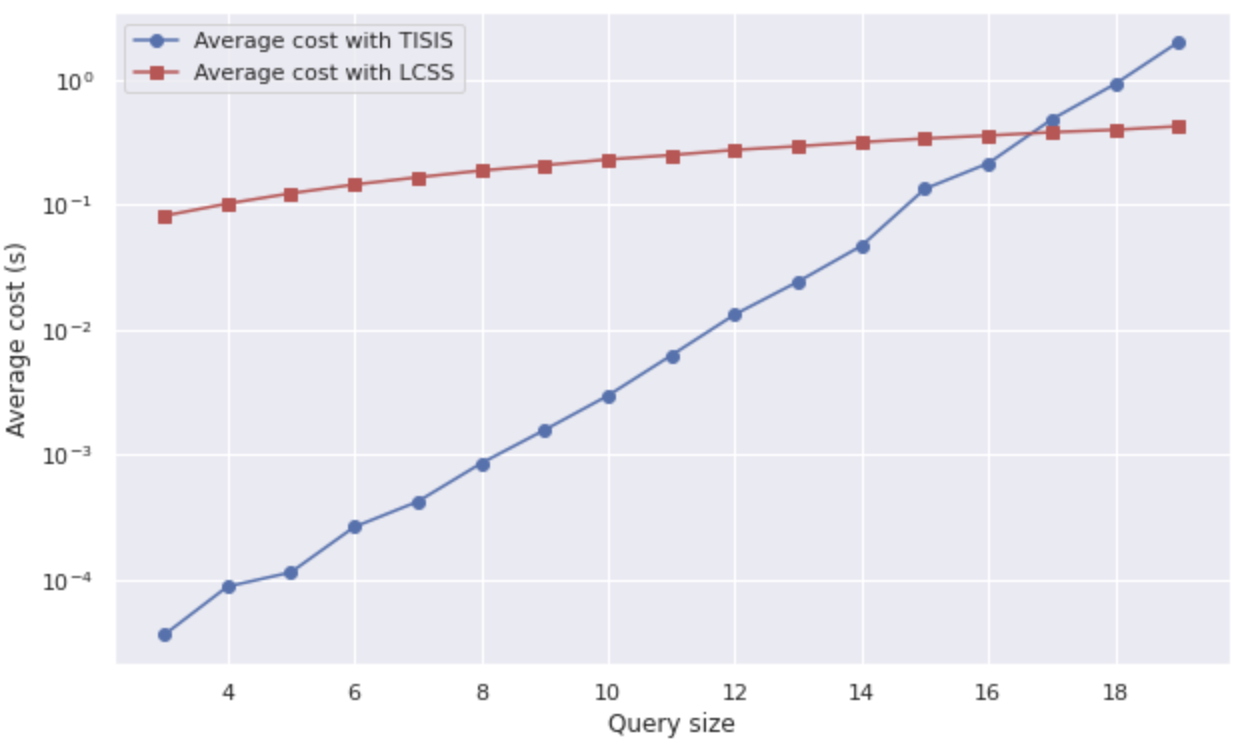}
    \caption{Average time as a function of query size}
    \label{fig:running_time_foursquare}
\end{figure}

\paragraph{Average computing time as a function of query size on Foursquare dataset, with a threshold of 0.5.}
\label{sec:Time as a function of query size}
In this section, we analyse the average execution time as a function of the query size specified by the user. To achieve this, we group our data by query size, and then calculate the average of all the time values for each size group. As previously noted, the threshold is fixed to 0.5 to represent the worst-case scenario for TISIS.
The results are illustrated in~\Cref{fig:tisis_small_sizes} and~\ref{fig:running_time_foursquare}. 

First, we observe on~\Cref{fig:tisis_small_sizes} that TISIS response time in less than 1 ms for query size smaller than 8. Thus, TISIS meets users expectation in terms of response time.

On ~\Cref{fig:running_time_foursquare}, we notice that our solution (TISIS in blue) is faster than LCSS-based baseline (in red) for all trajectories smaller than 17, \ie those containing less than 17 POIs, above which size it becomes slower than  LCSS-based baseline. 
However, we believe that it is not realistic to consider queries of a greater size (users would not reasonably ask queries with more POIs). 

Therefore, we can claim that TISIS largely outperforms the LCSS-based baseline for all realistic queries.

Note that we also experimented TISIS and  LCSS-based baseline for other threshold values (0.1 and 0.9). As expected, the results are more in favour of TISIS than for 0.5.

\begin{figure}
    \centering
    \includegraphics[width=\linewidth]{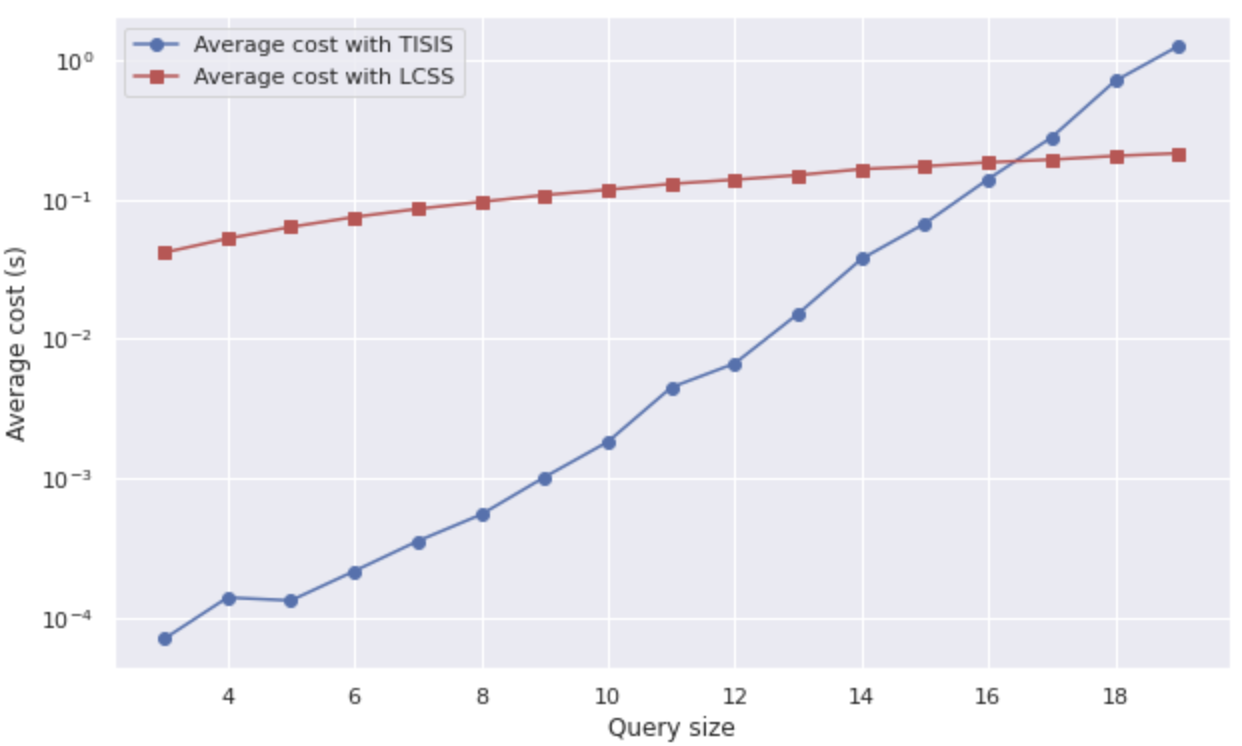}
    \caption{Average time as a function of query size for Gowalla dataset}
    \label{fig:gowalla}
\end{figure}

\begin{figure}
    \centering
    \includegraphics[width=\linewidth]{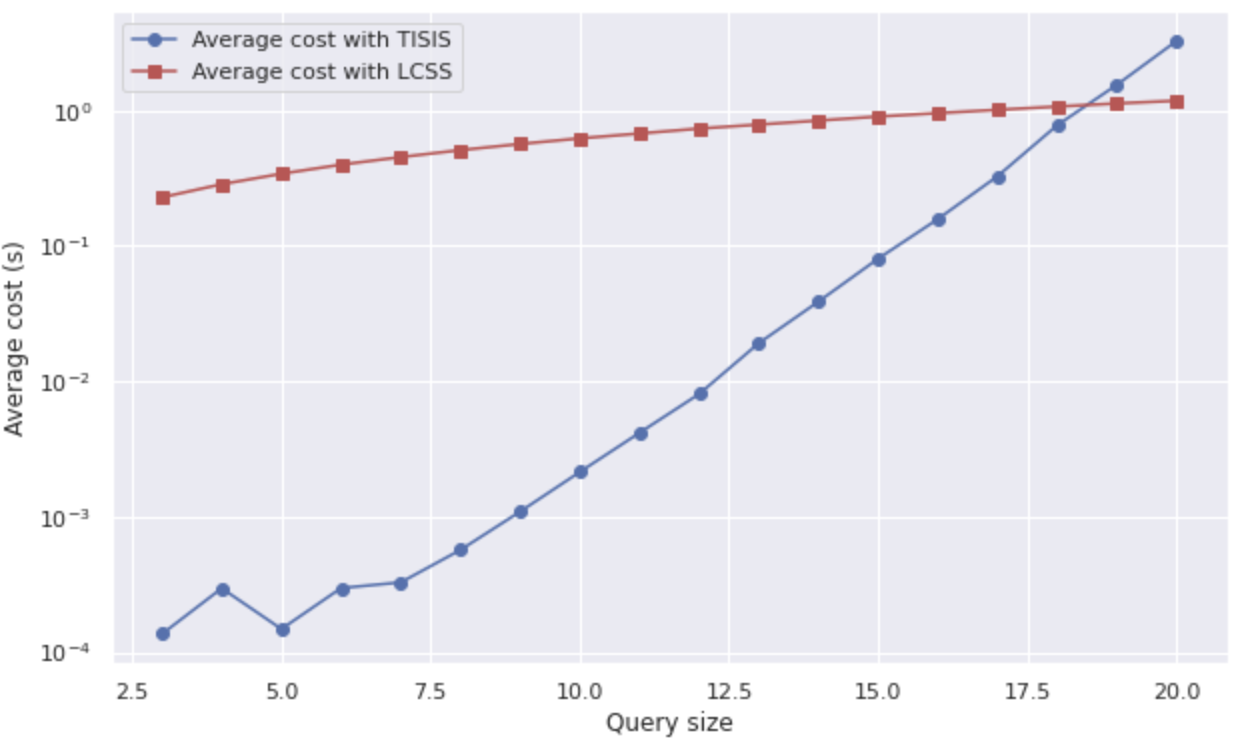}
    \caption{Average time as a function of query size for YFCC dataset}
    \label{fig:yfcc}
\end{figure}

\paragraph{Comparison of results with other datasets (YFCC, Gowalla).}
\label{sec:comparaison_datasets}
We performed the same experiment on other datasets to show that TISIS can produce good results on any other dataset. The chosen datasets are : Gowalla and YFCC .
The~\Cref{fig:gowalla} and~\ref{fig:yfcc} show the results obtained with the experiments conducted on these 2 datasets.

We observe that the results are comparable with the ones obtained for Foursquare, which confirms that TISIS outperforms LCSS-based baseline for reasonable query sizes and various datasets.
We highlight the benefit of TISIS for queries of average trajectory size (6 and 5 for Gowalla and YFCC respectively). TISIS is 330x (resp. 2200x) faster than  LCSS-based baseline.

\subsubsection{Exact indexing with 2 POIs results}
\label{sec:results_2POIs}

\begin{figure}
    \centering
    \includegraphics[width=\linewidth]{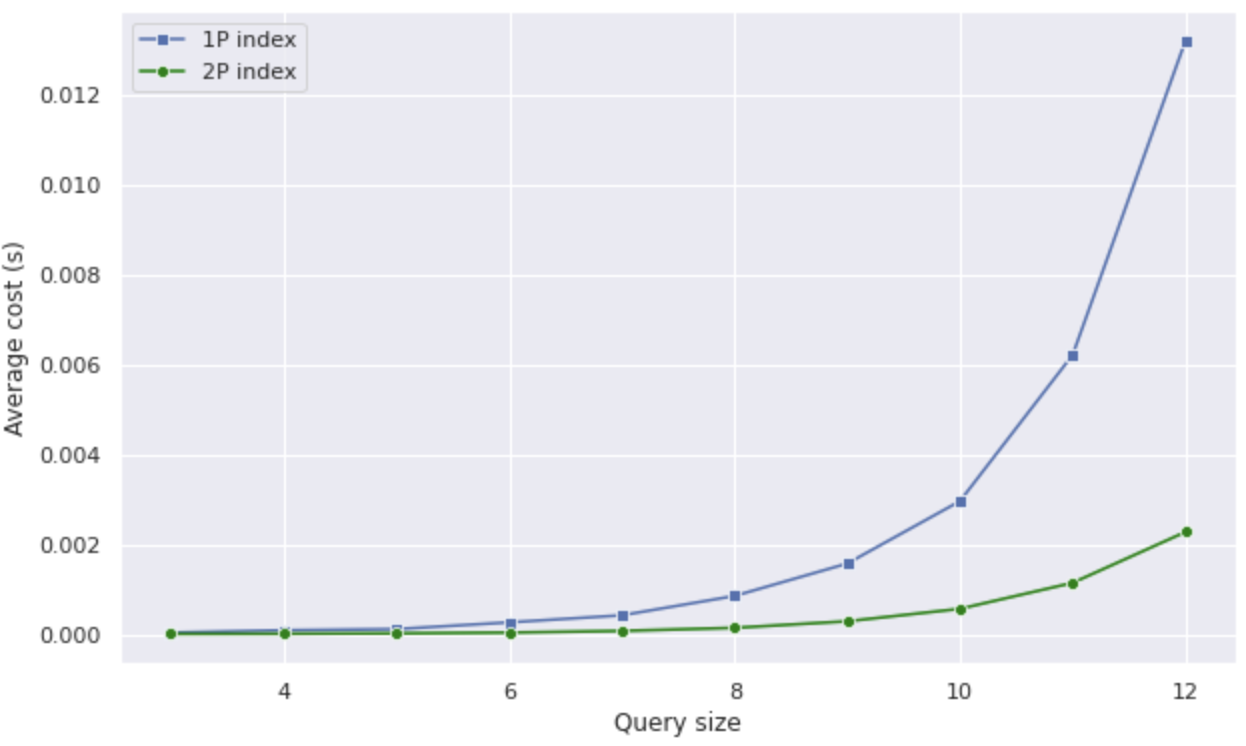}
    \caption{1 Point index vs. 2 Points index}
    \label{fig:comparaison1P_2P}
\end{figure}

\begin{figure}
    \centering
    \includegraphics[width=\linewidth]{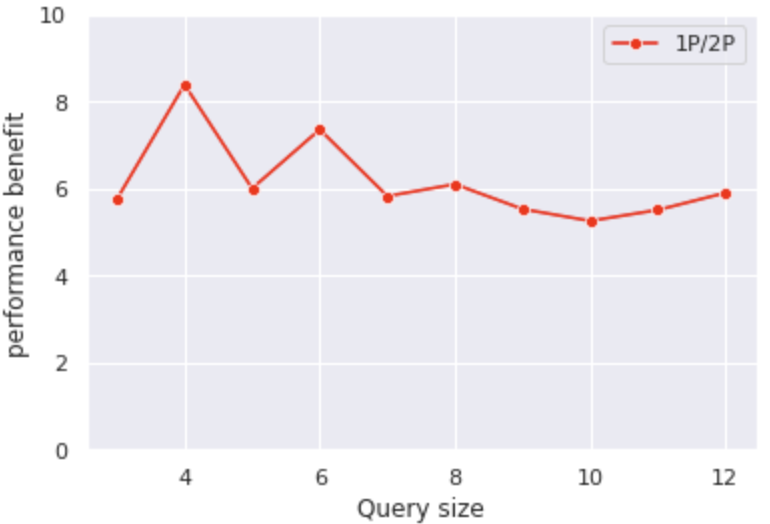}
    \caption{response time 1P/2P}
    \label{fig:response_time_1P_2P}
\end{figure}

We compare the performance of the single POI index (1P) and of the POI-pair index (2P) using the Foursquare dataset. 
As shown in \Cref{fig:comparaison1P_2P}, the POI-pair index (represented by the green curve) performs faster than the single POI index (represented by the blue curve).
On~\Cref{fig:response_time_1P_2P}, the performance benefit ranges from 5x to 8x (with an average benefit of 6x) for queries of size 3 to 12. 

\begin{table}
\centering
  \caption{Index construction cost}
  \label{fig:index_construct_cost}
  \begin{tabular}{ccl}
    \toprule
      &Index with 1P&Index with 2P\\
    \midrule
    \#entries & $2,900$ & $40,900$\\
    Avg \#trajectories& $15 \pm 20.24$ & $3 \pm 4.74$\\    
    Build time (ms)&$12$& $120$\\ 
  \bottomrule
\end{tabular}
\end{table}

As a counter part, the index creation time increases from 12ms to 120ms as reported on~\Cref{fig:index_construct_cost}. Indeed, the 2P index is 14x larger (40,900 entries) than the 1P index (2,900 entries). We can note that the number of entries in the 2P index remains tractable in practice.
As anticipated, the 2P index exhibits a relatively high degree of selectivity, granting access to only 3 trajectories out of 10K on average.

\subsubsection{Embedding-enhanced indexing-based results }
\label{sec:results_embeddings}
\begin{figure}
    \centering
    \includegraphics[width=\linewidth]{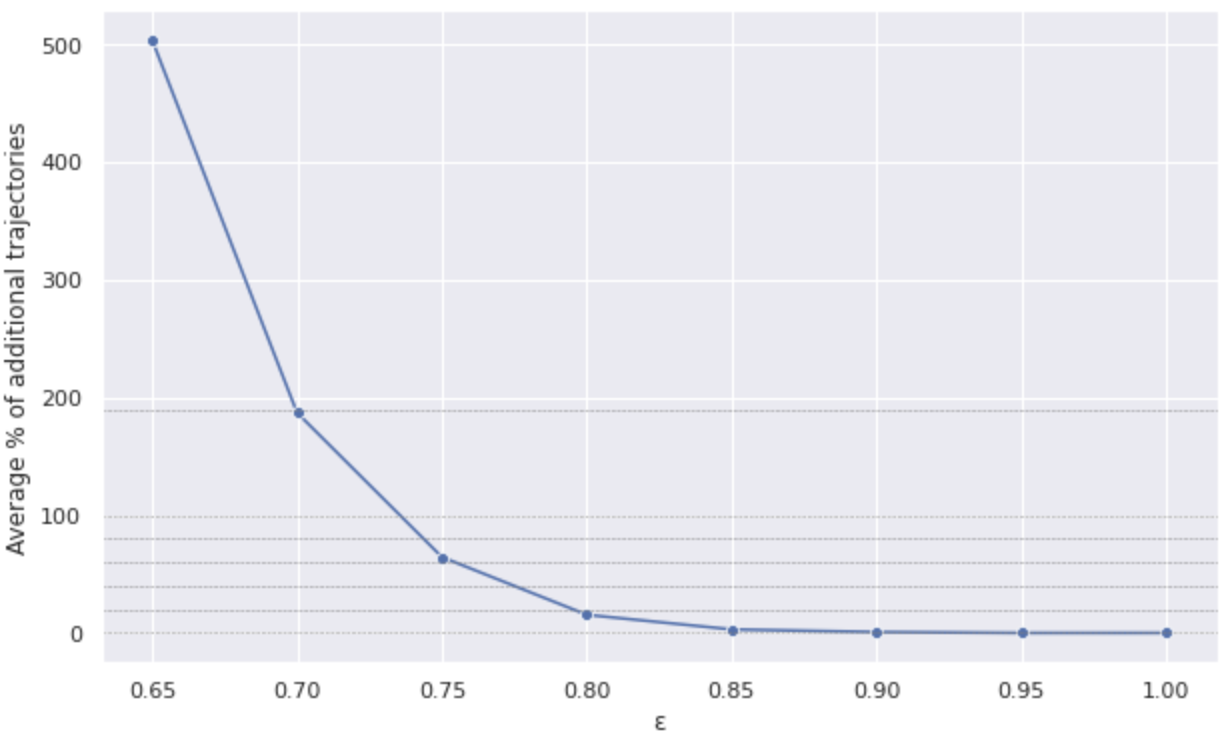}
    \caption{Effect of $\epsilon$ on the average percentage of additional trajectories }
    \label{figure:pourcentage_traj_supp}
\end{figure}

We report the number of trajectories on \Cref{figure:pourcentage_traj_supp}. We vary $\epsilon$ on the x-axis. The y-axis is the average percentage of extra trajectories returned by the queries. For example, for $\epsilon=0.72$ we get twice as many trajectories.
This increase is due to the fact that the number of neighbors increases with~$\epsilon$. Specifically, we obtained 14 neighbors for $\epsilon=0.72$ and only 1 neighbor for $\epsilon\ge 0.9$.

Moreover, the cost remains close to that of exact TISIS: less than 0.7s for $\epsilon>0.72$, demonstrating the efficiency of our approach.
For $\epsilon<0.72$ the cost actually increases significantly (up to 1.4s for $\epsilon=0.65$), 
but such a range of $\epsilon$ values would not be used in practice because it returns too many trajectories (more than twice as many as with exact TISIS).

Several recent works have demonstrated the efficacy of using embeddings to represent mobility trajectories or POIs for various tasks, such as location classification and next location prediction. Notable examples include~\cite{Embedding_Enhanced_Similarity_Metrics_for_Next_POI_Recommendation,location_embeddings_from_trajectories1,location_embeddings_from_trajectories2,location_embeddings_from_trajectories3}

\begin{figure}[H]
    \centering
    \includegraphics[width=\linewidth]{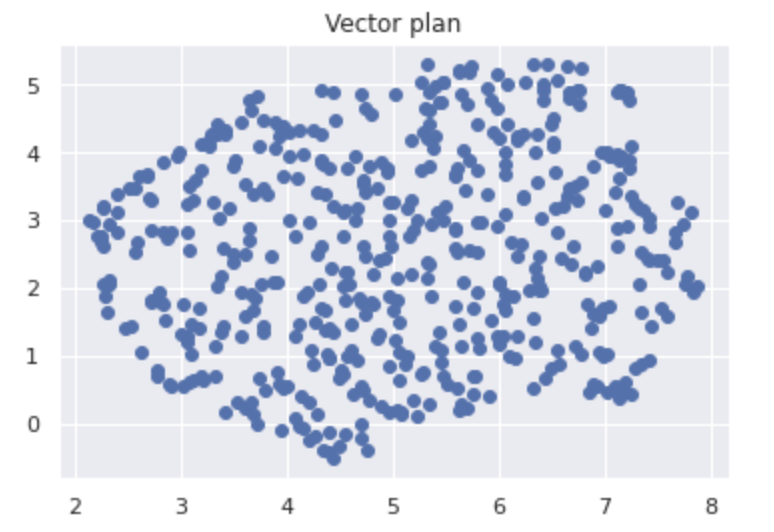}
    \caption{500 POIs sample (in 2D plan)}
    \label{fig:distribution_POIs_foursquare}
\end{figure}

One way to assess the relevance of the embeddings is to verify that they are well dispersed on the vector plan, which is illustrated in (\Cref{fig:distribution_POIs_foursquare}).

\begin{figure}[H]
    \centering
    \includegraphics[width=\linewidth]{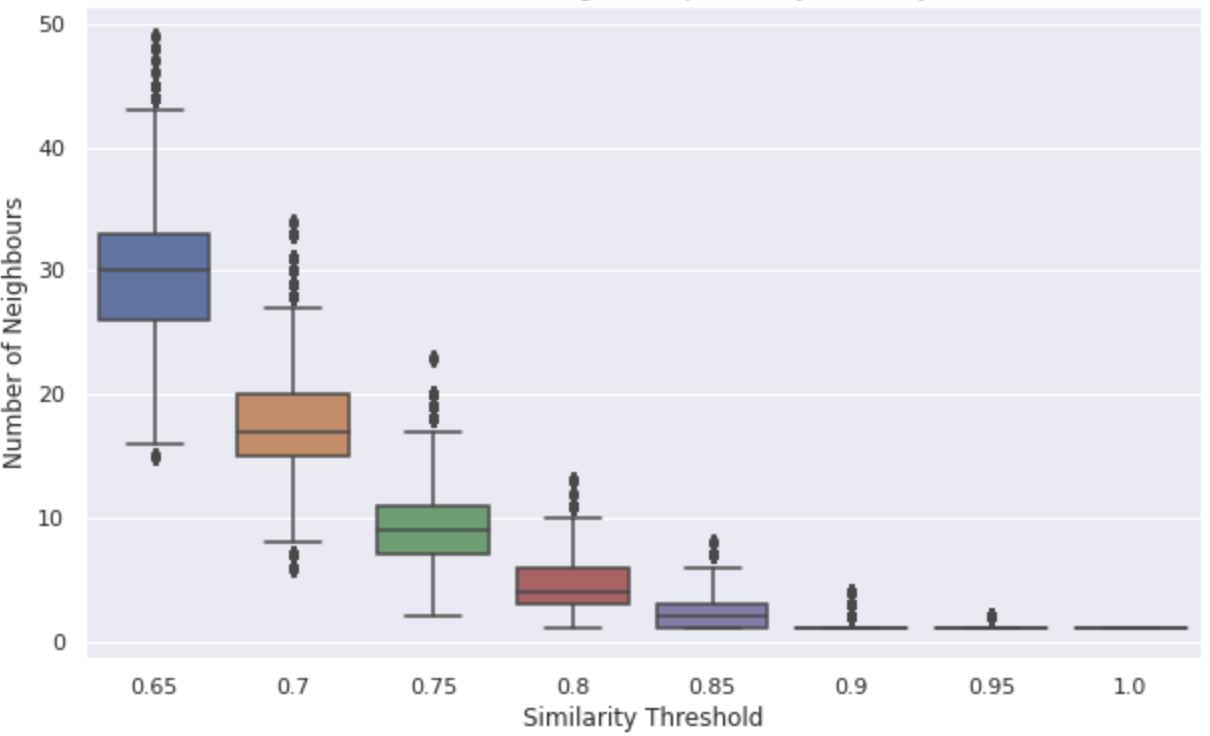}
    \caption{\#Neighbours per POI }
    \label{fig:voisins_fct_seuils}
\end{figure}

Another approach to assess the relevance of embeddings is to show that the number of neighbours for any POI is increasing gradually with the similarity threshold.
This is the case on (\Cref{fig:voisins_fct_seuils})
where we plot the number of neighbours per POI according to the threshold 
(boxplots in this figure highlight the percentiles 5,25,50,75,95 of the number of neighbours).

\section{Conclusion}
\label{sec:conclusion}
This paper demonstrates the effectiveness of an indexing-based approach for trajectory similarity search when compared to a baseline method that relies on the Longest Common Subsequence (LCSS) algorithm.
To address the performance limitations inherent in LCSS-based search, we introduce \textit{TISIS} (Trajectory Indexing for Similarity Search). TISIS is an efficient method that employs an exact indexing approach for trajectory similarity search. By leveraging two distinct trajectory indexing techniques, our approach consistently produces identical results to the LCSS-based baseline, while offering significantly improved performance.

Nevertheless, TISIS may be overly restrictive, as it relies on exact matching of POIs, which can result in too few (or even no) returned results. To mitigate this limitation, we have developed and implemented \textit{TISIS*}, a relaxed variant of TISIS. 
This extension enables more comprehensive retrieval of similar trajectories by accounting for semantic similarities between POIs, rather than relying solely on exact matches. The relaxed method yields a significantly higher number of results (approximately twice as many as the exact method with a similarity threshold of 0.72) while preserving computational efficiency.

We conducted experiments on three real-world datasets, demonstrating that the proposed approach significantly outperforms the LCSS-based baseline and is adaptable to various datasets.

Currently, \textit{TISIS} is limited to optimize the LCSS-based similarity measure, as it retrieves trajectories that share a specific number of POIs with the query, in the same sequential order.

As part of our future work, we aim to explore the problem of retrieving the top $K$ trajectories with the highest similarity to a given query. This will require assigning a similarity score to each trajectory.
For the exact matching approach, a potential scoring method could involve counting the number of subsequences in a trajectory that match subsequences in the query. However, for the embedding-based approach, the scoring process becomes complex, as it must account for the contextual similarity between POIs in addition to exact matches.

\bibliographystyle{ACM-Reference-Format}
\bibliography{bib}


\begin{thebibliography}{31}


\ifx \showCODEN    \undefined \def \showCODEN     #1{\unskip}     \fi
\ifx \showDOI      \undefined \def \showDOI       #1{#1}\fi
\ifx \showISBNx    \undefined \def \showISBNx     #1{\unskip}     \fi
\ifx \showISBNxiii \undefined \def \showISBNxiii  #1{\unskip}     \fi
\ifx \showISSN     \undefined \def \showISSN      #1{\unskip}     \fi
\ifx \showLCCN     \undefined \def \showLCCN      #1{\unskip}     \fi
\ifx \shownote     \undefined \def \shownote      #1{#1}          \fi
\ifx \showarticletitle \undefined \def \showarticletitle #1{#1}   \fi
\ifx \showURL      \undefined \def \showURL       {\relax}        \fi
\providecommand\bibfield[2]{#2}
\providecommand\bibinfo[2]{#2}
\providecommand\natexlab[1]{#1}
\providecommand\showeprint[2][]{arXiv:#2}

\bibitem[Altschul et~al\mbox{.}(1990)]%
        {Blast}
\bibfield{author}{\bibinfo{person}{Stephen~F Altschul}, \bibinfo{person}{Warren Gish}, \bibinfo{person}{Webb Miller}, \bibinfo{person}{Eugene~W Myers}, {and} \bibinfo{person}{David~J Lipman}.} \bibinfo{year}{1990}\natexlab{}.
\newblock \showarticletitle{Basic local alignment search tool}.
\newblock \bibinfo{journal}{\emph{Journal of molecular biology}} \bibinfo{volume}{215}, \bibinfo{number}{3} (\bibinfo{year}{1990}), \bibinfo{pages}{403--410}.
\newblock


\bibitem[Berndt and Clifford(1994)]%
        {DTW1}
\bibfield{author}{\bibinfo{person}{Donald~J Berndt} {and} \bibinfo{person}{James Clifford}.} \bibinfo{year}{1994}\natexlab{}.
\newblock \showarticletitle{Using dynamic time warping to find patterns in time series}. In \bibinfo{booktitle}{\emph{Proceedings of the 3rd international conference on knowledge discovery and data mining}}. \bibinfo{pages}{359--370}.
\newblock


\bibitem[Biester et~al\mbox{.}(2020)]%
        {location_embeddings_from_trajectories2}
\bibfield{author}{\bibinfo{person}{Laura Biester}, \bibinfo{person}{Carmen Banea}, {and} \bibinfo{person}{Rada Mihalcea}.} \bibinfo{year}{2020}\natexlab{}.
\newblock \showarticletitle{Building Location Embeddings from Physical Trajectories and Textual Representations}. In \bibinfo{booktitle}{\emph{Proceedings of the 1st Conference of the Asia-Pacific Chapter of the Association for Computational Linguistics and the 10th International Joint Conference on Natural Language Processing}}, \bibfield{editor}{\bibinfo{person}{Kam-Fai Wong}, \bibinfo{person}{Kevin Knight}, {and} \bibinfo{person}{Hua Wu}} (Eds.). \bibinfo{publisher}{Association for Computational Linguistics}, \bibinfo{address}{Suzhou, China}, \bibinfo{pages}{425--434}.
\newblock
\urldef\tempurl%
\url{https://aclanthology.org/2020.aacl-main.44}
\showURL{%
\tempurl}


\bibitem[Bringmann et~al\mbox{.}(2023)]%
        {approximation_LCSS2}
\bibfield{author}{\bibinfo{person}{Karl Bringmann}, \bibinfo{person}{Vincent Cohen-Addad}, {and} \bibinfo{person}{Debarati Das}.} \bibinfo{year}{2023}\natexlab{}.
\newblock \showarticletitle{A Linear-Time n0.4-Approximation for Longest Common Subsequence}.
\newblock \bibinfo{journal}{\emph{ACM Trans. Algorithms}} \bibinfo{volume}{19}, \bibinfo{number}{1}, Article \bibinfo{articleno}{9} (\bibinfo{date}{feb} \bibinfo{year}{2023}), \bibinfo{numpages}{24}~pages.
\newblock
\showISSN{1549-6325}
\urldef\tempurl%
\url{https://doi.org/10.1145/3568398}
\showDOI{\tempurl}


\bibitem[Chan et~al\mbox{.}(2007)]%
        {LCIS}
\bibfield{author}{\bibinfo{person}{Wun{-}Tat Chan}, \bibinfo{person}{Yong Zhang}, \bibinfo{person}{Stanley P.~Y. Fung}, \bibinfo{person}{Deshi Ye}, {and} \bibinfo{person}{Hong Zhu}.} \bibinfo{year}{2007}\natexlab{}.
\newblock \showarticletitle{Efficient algorithms for finding a longest common increasing subsequence}.
\newblock \bibinfo{journal}{\emph{J. Comb. im.}} \bibinfo{volume}{13}, \bibinfo{number}{3} (\bibinfo{year}{2007}), \bibinfo{pages}{277--288}.
\newblock
\urldef\tempurl%
\url{https://doi.org/10.1007/S10878-006-9031-7}
\showDOI{\tempurl}


\bibitem[Chen and Ng(2004)]%
        {ERP}
\bibfield{author}{\bibinfo{person}{Lei Chen} {and} \bibinfo{person}{Raymond Ng}.} \bibinfo{year}{2004}\natexlab{}.
\newblock \showarticletitle{On The Marriage of Lp-norms and Edit Distance.} \bibinfo{pages}{792--803}.
\newblock


\bibitem[Chen et~al\mbox{.}(2005)]%
        {EDR}
\bibfield{author}{\bibinfo{person}{Lei Chen}, \bibinfo{person}{M.~Tamer \"{O}zsu}, {and} \bibinfo{person}{Vincent Oria}.} \bibinfo{year}{2005}\natexlab{}.
\newblock \showarticletitle{Robust and fast similarity search for moving object trajectories}. In \bibinfo{booktitle}{\emph{Proceedings of the 2005 ACM SIGMOD International Conference on Management of Data}} (Baltimore, Maryland) \emph{(\bibinfo{series}{SIGMOD '05})}. \bibinfo{publisher}{Association for Computing Machinery}, \bibinfo{address}{New York, NY, USA}, \bibinfo{pages}{491–502}.
\newblock
\showISBNx{1595930604}
\urldef\tempurl%
\url{https://doi.org/10.1145/1066157.1066213}
\showDOI{\tempurl}


\bibitem[Cho et~al\mbox{.}(2011)]%
        {gowalla}
\bibfield{author}{\bibinfo{person}{Eunjoon Cho}, \bibinfo{person}{Seth~A. Myers}, {and} \bibinfo{person}{Jure Leskovec}.} \bibinfo{year}{2011}\natexlab{}.
\newblock \showarticletitle{Friendship and mobility: user movement in location-based social networks}. In \bibinfo{booktitle}{\emph{Proceedings of the 17th {ACM} {SIGKDD} International Conference on Knowledge Discovery and Data Mining, San Diego, CA, USA, 2011}}, \bibfield{editor}{\bibinfo{person}{Chid Apt{\'{e}}}, \bibinfo{person}{Joydeep Ghosh}, {and} \bibinfo{person}{Padhraic Smyth}} (Eds.). \bibinfo{publisher}{{ACM}}, \bibinfo{pages}{1082--1090}.
\newblock
\urldef\tempurl%
\url{https://doi.org/10.1145/2020408.2020579}
\showDOI{\tempurl}


\bibitem[Gotoh(1982)]%
        {alignement_local2}
\bibfield{author}{\bibinfo{person}{Osamu Gotoh}.} \bibinfo{year}{1982}\natexlab{}.
\newblock \showarticletitle{An improved algorithm for matching biological sequences}.
\newblock \bibinfo{journal}{\emph{Journal of molecular biology}} \bibinfo{volume}{162}, \bibinfo{number}{3} (\bibinfo{year}{1982}), \bibinfo{pages}{705--708}.
\newblock


\bibitem[Hajiaghayi et~al\mbox{.}(2020)]%
        {approximation_LCSS1}
\bibfield{author}{\bibinfo{person}{MohammadTaghi Hajiaghayi}, \bibinfo{person}{Masoud Seddighin}, \bibinfo{person}{Saeed Seddighin}, {and} \bibinfo{person}{Xiaorui Sun}.} \bibinfo{year}{2020}\natexlab{}.
\newblock \showarticletitle{Approximating {LCS} in Linear Time: Beating the {\(\surd\)}n Barrier}.
\newblock \bibinfo{journal}{\emph{CoRR}}  \bibinfo{volume}{abs/2003.07285} (\bibinfo{year}{2020}).
\newblock
\showeprint[arXiv]{2003.07285}
\urldef\tempurl%
\url{https://arxiv.org/abs/2003.07285}
\showURL{%
\tempurl}


\bibitem[Han et~al\mbox{.}(2007)]%
        {LCS_timeSeries}
\bibfield{author}{\bibinfo{person}{Tae~Sik Han}, \bibinfo{person}{Seung{-}Kyu Ko}, {and} \bibinfo{person}{Jaewoo Kang}.} \bibinfo{year}{2007}\natexlab{}.
\newblock \showarticletitle{Efficient Subsequence Matching Using the Longest Common Subsequence with a Dual Match Index}. In \bibinfo{booktitle}{\emph{Machine Learning and Data Mining in Pattern Recognition, 5th International Conference, {MLDM} 2007, Leipzig, Germany, July 18-20, 2007, Proceedings}} \emph{(\bibinfo{series}{Lecture Notes in Computer Science}, Vol.~\bibinfo{volume}{4571})}, \bibfield{editor}{\bibinfo{person}{Petra Perner}} (Ed.). \bibinfo{publisher}{Springer}, \bibinfo{pages}{585--600}.
\newblock
\urldef\tempurl%
\url{https://doi.org/10.1007/978-3-540-73499-4\_44}
\showDOI{\tempurl}


\bibitem[Hirschberg(1975)]%
        {lcss_classique1}
\bibfield{author}{\bibinfo{person}{D.~S. Hirschberg}.} \bibinfo{year}{1975}\natexlab{}.
\newblock \showarticletitle{A linear space algorithm for computing maximal common subsequences}.
\newblock \bibinfo{journal}{\emph{Commun. ACM}} \bibinfo{volume}{18}, \bibinfo{number}{6} (\bibinfo{date}{jun} \bibinfo{year}{1975}), \bibinfo{pages}{341–343}.
\newblock
\showISSN{0001-0782}
\urldef\tempurl%
\url{https://doi.org/10.1145/360825.360861}
\showDOI{\tempurl}


\bibitem[Hirschberg(1977)]%
        {lcss2}
\bibfield{author}{\bibinfo{person}{Daniel~S. Hirschberg}.} \bibinfo{year}{1977}\natexlab{}.
\newblock \showarticletitle{Algorithms for the Longest Common Subsequence Problem}.
\newblock \bibinfo{journal}{\emph{J. ACM}} \bibinfo{volume}{24}, \bibinfo{number}{4} (\bibinfo{date}{oct} \bibinfo{year}{1977}), \bibinfo{pages}{664–675}.
\newblock
\showISSN{0004-5411}
\urldef\tempurl%
\url{https://doi.org/10.1145/322033.322044}
\showDOI{\tempurl}


\bibitem[Jarrad et~al\mbox{.}(2023)]%
        {Embedding_Enhanced_Similarity_Metrics_for_Next_POI_Recommendation}
\bibfield{author}{\bibinfo{person}{Sara Jarrad}, \bibinfo{person}{Hubert Naacke}, \bibinfo{person}{St{\'{e}}phane Gan{\c{c}}arski}, {and} \bibinfo{person}{Modou Gueye}.} \bibinfo{year}{2023}\natexlab{}.
\newblock \showarticletitle{Embedding-Enhanced Similarity Metrics for Next {POI} Recommendation}. In \bibinfo{booktitle}{\emph{Proceedings of the 12th International Conference on Data Science, Technology and Applications, {DATA} 2023, Rome, Italy, July 11-13, 2023}}, \bibfield{editor}{\bibinfo{person}{Oleg Gusikhin}, \bibinfo{person}{Slimane Hammoudi}, {and} \bibinfo{person}{Alfredo Cuzzocrea}} (Eds.). \bibinfo{publisher}{{SCITEPRESS}}, \bibinfo{pages}{247--254}.
\newblock
\urldef\tempurl%
\url{https://doi.org/10.5220/0012060300003541}
\showDOI{\tempurl}


\bibitem[Lin et~al\mbox{.}(2021)]%
        {location_embeddings_from_trajectories1}
\bibfield{author}{\bibinfo{person}{Yan Lin}, \bibinfo{person}{Huaiyu Wan}, \bibinfo{person}{Shengnan Guo}, {and} \bibinfo{person}{Youfang Lin}.} \bibinfo{year}{2021}\natexlab{}.
\newblock \showarticletitle{Pre-training Context and Time Aware Location Embeddings from Spatial-Temporal Trajectories for User Next Location Prediction}.
\newblock \bibinfo{journal}{\emph{Proceedings of the AAAI Conference on Artificial Intelligence}} \bibinfo{volume}{35}, \bibinfo{number}{5} (\bibinfo{date}{May} \bibinfo{year}{2021}), \bibinfo{pages}{4241--4248}.
\newblock
\urldef\tempurl%
\url{https://doi.org/10.1609/aaai.v35i5.16548}
\showDOI{\tempurl}


\bibitem[Magdy et~al\mbox{.}(2015)]%
        {Review_on_trajectory_similarity_measures}
\bibfield{author}{\bibinfo{person}{Nehal Magdy}, \bibinfo{person}{Mahmoud Sakr}, \bibinfo{person}{Tamer Abdelkader}, {and} \bibinfo{person}{Khaled Elbahnasy}.} \bibinfo{year}{2015}\natexlab{}.
\newblock \showarticletitle{Review on trajectory similarity measures}.
\newblock
\urldef\tempurl%
\url{https://doi.org/10.1109/IntelCIS.2015.7397286}
\showDOI{\tempurl}


\bibitem[Mikolov et~al\mbox{.}(2013)]%
        {efficient_word_estimation_using_w2v}
\bibfield{author}{\bibinfo{person}{Tom{\'{a}}s Mikolov}, \bibinfo{person}{Kai Chen}, \bibinfo{person}{Greg Corrado}, {and} \bibinfo{person}{Jeffrey Dean}.} \bibinfo{year}{2013}\natexlab{}.
\newblock \showarticletitle{Efficient Estimation of Word Representations in Vector Space}. In \bibinfo{booktitle}{\emph{1st International Conference on Learning Representations, {ICLR} 2013, Scottsdale, Arizona, USA, May 2-4, 2013, Workshop Track Proceedings}}, \bibfield{editor}{\bibinfo{person}{Yoshua Bengio} {and} \bibinfo{person}{Yann LeCun}} (Eds.).
\newblock
\urldef\tempurl%
\url{http://arxiv.org/abs/1301.3781}
\showURL{%
\tempurl}


\bibitem[Needleman and Wunsch(1970)]%
        {alignement_global1}
\bibfield{author}{\bibinfo{person}{Saul~B. Needleman} {and} \bibinfo{person}{Christian~D. Wunsch}.} \bibinfo{year}{1970}\natexlab{}.
\newblock \showarticletitle{A general method applicable to the search for similarities in the amino acid sequence of two proteins}.
\newblock \bibinfo{journal}{\emph{Journal of Molecular Biology}} \bibinfo{volume}{48}, \bibinfo{number}{3} (\bibinfo{year}{1970}), \bibinfo{pages}{443--453}.
\newblock
\showISSN{0022-2836}
\urldef\tempurl%
\url{https://doi.org/10.1016/0022-2836(70)90057-4}
\showDOI{\tempurl}


\bibitem[Notredame et~al\mbox{.}(2000)]%
        {alignement_multiple}
\bibfield{author}{\bibinfo{person}{C{\'e}dric Notredame}, \bibinfo{person}{Desmond~G Higgins}, {and} \bibinfo{person}{Jaap Heringa}.} \bibinfo{year}{2000}\natexlab{}.
\newblock \showarticletitle{T-Coffee: A novel method for fast and accurate multiple sequence alignment}.
\newblock \bibinfo{journal}{\emph{Journal of molecular biology}} \bibinfo{volume}{302}, \bibinfo{number}{1} (\bibinfo{year}{2000}), \bibinfo{pages}{205--217}.
\newblock


\bibitem[Sellers(1974)]%
        {lcss_classique3}
\bibfield{author}{\bibinfo{person}{Peter~H Sellers}.} \bibinfo{year}{1974}\natexlab{}.
\newblock \showarticletitle{An algorithm for the distance between two finite sequences}.
\newblock \bibinfo{journal}{\emph{Journal of Combinatorial Theory}} \bibinfo{volume}{16}, \bibinfo{number}{2} (\bibinfo{year}{1974}), \bibinfo{pages}{253--258}.
\newblock
\showISSN{0097-3165}
\urldef\tempurl%
\url{https://doi.org/10.1016/0097-3165(74)90050-8}
\showDOI{\tempurl}


\bibitem[Smith et~al\mbox{.}(1981a)]%
        {alignement_local1}
\bibfield{author}{\bibinfo{person}{Temple~F Smith}, \bibinfo{person}{Michael~S Waterman}, {et~al\mbox{.}}} \bibinfo{year}{1981}\natexlab{a}.
\newblock \showarticletitle{Identification of common molecular subsequences}.
\newblock \bibinfo{journal}{\emph{Journal of molecular biology}} \bibinfo{volume}{147}, \bibinfo{number}{1} (\bibinfo{year}{1981}), \bibinfo{pages}{195--197}.
\newblock


\bibitem[Smith et~al\mbox{.}(1981b)]%
        {alignement_global2}
\bibfield{author}{\bibinfo{person}{Temple~F Smith}, \bibinfo{person}{Michael~S Waterman}, {et~al\mbox{.}}} \bibinfo{year}{1981}\natexlab{b}.
\newblock \showarticletitle{Identification of common molecular subsequences}.
\newblock \bibinfo{journal}{\emph{Journal of molecular biology}} \bibinfo{volume}{147}, \bibinfo{number}{1} (\bibinfo{year}{1981}), \bibinfo{pages}{195--197}.
\newblock


\bibitem[Su et~al\mbox{.}(2020)]%
        {survey_of_trajectory_distance_measures}
\bibfield{author}{\bibinfo{person}{Han Su}, \bibinfo{person}{Shuncheng Liu}, \bibinfo{person}{Bolong Zheng}, \bibinfo{person}{Xiaofang Zhou}, {and} \bibinfo{person}{Kai Zheng}.} \bibinfo{year}{2020}\natexlab{}.
\newblock \showarticletitle{A survey of trajectory distance measures and performance evaluation}.
\newblock \bibinfo{journal}{\emph{{VLDB} J.}} \bibinfo{volume}{29}, \bibinfo{number}{1} (\bibinfo{year}{2020}), \bibinfo{pages}{3--32}.
\newblock
\urldef\tempurl%
\url{https://doi.org/10.1007/S00778-019-00574-9}
\showDOI{\tempurl}


\bibitem[Thomee et~al\mbox{.}(2016)]%
        {yfcc}
\bibfield{author}{\bibinfo{person}{Bart Thomee}, \bibinfo{person}{David~A. Shamma}, \bibinfo{person}{Gerald Friedland}, \bibinfo{person}{Benjamin Elizalde}, \bibinfo{person}{Karl Ni}, \bibinfo{person}{Douglas Poland}, \bibinfo{person}{Damian Borth}, {and} \bibinfo{person}{Li{-}Jia Li}.} \bibinfo{year}{2016}\natexlab{}.
\newblock \showarticletitle{{YFCC100M:} the new data in multimedia research}.
\newblock \bibinfo{journal}{\emph{Commununications of the {ACM}}} \bibinfo{volume}{59}, \bibinfo{number}{2} (\bibinfo{year}{2016}), \bibinfo{pages}{64--73}.
\newblock


\bibitem[Toohey and Duckham(2015)]%
        {Trajectory_similarity_measures}
\bibfield{author}{\bibinfo{person}{Kevin Toohey} {and} \bibinfo{person}{Matt Duckham}.} \bibinfo{year}{2015}\natexlab{}.
\newblock \showarticletitle{Trajectory similarity measures}.
\newblock \bibinfo{journal}{\emph{{ACM} {SIGSPATIAL}}} \bibinfo{volume}{7}, \bibinfo{number}{1} (\bibinfo{year}{2015}), \bibinfo{pages}{43--50}.
\newblock
\urldef\tempurl%
\url{https://doi.org/10.1145/2782759.2782767}
\showDOI{\tempurl}


\bibitem[Vlachos et~al\mbox{.}(2002)]%
        {Discovering_Similar_Multidimensional_Trajectories}
\bibfield{author}{\bibinfo{person}{Michail Vlachos}, \bibinfo{person}{Dimitrios Gunopulos}, {and} \bibinfo{person}{George Kollios}.} \bibinfo{year}{2002}\natexlab{}.
\newblock \showarticletitle{Discovering Similar Multidimensional Trajectories}. In \bibinfo{booktitle}{\emph{Proceedings of the 18th International Conference on Data Engineering}}, \bibfield{editor}{\bibinfo{person}{Rakesh Agrawal} {and} \bibinfo{person}{Klaus~R. Dittrich}} (Eds.). \bibinfo{publisher}{{IEEE} Computer Society}, \bibinfo{pages}{673--684}.
\newblock
\urldef\tempurl%
\url{https://doi.org/10.1109/ICDE.2002.994784}
\showDOI{\tempurl}


\bibitem[Vlachos et~al\mbox{.}(2003)]%
        {Indexing_multi_dimensional_time_series}
\bibfield{author}{\bibinfo{person}{Michail Vlachos}, \bibinfo{person}{Marios Hadjieleftheriou}, \bibinfo{person}{Dimitrios Gunopulos}, {and} \bibinfo{person}{Eamonn~J. Keogh}.} \bibinfo{year}{2003}\natexlab{}.
\newblock \showarticletitle{Indexing multi-dimensional time-series with support for multiple distance measures}. In \bibinfo{booktitle}{\emph{Proceedings of the Ninth {ACM} {SIGKDD} International Conference on Knowledge Discovery and Data Mining, Washington, DC, USA, 2003}}, \bibfield{editor}{\bibinfo{person}{Lise Getoor}, \bibinfo{person}{Ted~E. Senator}, \bibinfo{person}{Pedro~M. Domingos}, {and} \bibinfo{person}{Christos Faloutsos}} (Eds.). \bibinfo{publisher}{{ACM}}, \bibinfo{pages}{216--225}.
\newblock
\urldef\tempurl%
\url{https://doi.org/10.1145/956750.956777}
\showDOI{\tempurl}


\bibitem[Wan et~al\mbox{.}(2022)]%
        {location_embeddings_from_trajectories3}
\bibfield{author}{\bibinfo{person}{Huaiyu Wan}, \bibinfo{person}{Yan Lin}, \bibinfo{person}{Shengnan Guo}, {and} \bibinfo{person}{Youfang Lin}.} \bibinfo{year}{2022}\natexlab{}.
\newblock \showarticletitle{Pre-Training Time-Aware Location Embeddings from Spatial-Temporal Trajectories}.
\newblock \bibinfo{journal}{\emph{IEEE Transactions on Knowledge and Data Engineering}} \bibinfo{volume}{34}, \bibinfo{number}{11} (\bibinfo{year}{2022}), \bibinfo{pages}{5510--5523}.
\newblock
\urldef\tempurl%
\url{https://doi.org/10.1109/TKDE.2021.3057875}
\showDOI{\tempurl}


\bibitem[Wang et~al\mbox{.}(2013)]%
        {An_effectiveness_study_on_trajectory_similarity_measures}
\bibfield{author}{\bibinfo{person}{Haozhou Wang}, \bibinfo{person}{Han Su}, \bibinfo{person}{Kai Zheng}, \bibinfo{person}{Shazia Sadiq}, {and} \bibinfo{person}{Xiaofang Zhou}.} \bibinfo{year}{2013}\natexlab{}.
\newblock \showarticletitle{An effectiveness study on trajectory similarity measures}. In \bibinfo{booktitle}{\emph{Proceedings of the Twenty-Fourth Australasian Database Conference}} (Adelaide, Australia) \emph{(\bibinfo{series}{ADC '13})}. \bibinfo{publisher}{Australian Computer Society, Inc.}, \bibinfo{address}{AUS}, \bibinfo{pages}{13–22}.
\newblock
\showISBNx{9781921770227}


\bibitem[Yang et~al\mbox{.}(2015)]%
        {foursquare_dataset}
\bibfield{author}{\bibinfo{person}{Dingqi Yang}, \bibinfo{person}{Daqing Zhang}, \bibinfo{person}{Vincent.~W. Zheng}, {and} \bibinfo{person}{Zhiyong Yu}.} \bibinfo{year}{2015}\natexlab{}.
\newblock \showarticletitle{Modeling User Activity Preference by Leveraging User Spatial Temporal Characteristics in LBSNs}.
\newblock \bibinfo{journal}{\emph{IEEE Transactions on Systems, Man, and Cybernetics: Systems}} \bibinfo{volume}{45}, \bibinfo{number}{1} (\bibinfo{year}{2015}), \bibinfo{pages}{129--142}.
\newblock
\showISSN{2168-2216}


\bibitem[Yi et~al\mbox{.}(1999)]%
        {DTW2}
\bibfield{author}{\bibinfo{person}{Byoung-Kee Yi}, \bibinfo{person}{H. Jagadish}, {and} \bibinfo{person}{Christos Faloutsos}.} \bibinfo{year}{1999}\natexlab{}.
\newblock \showarticletitle{Efficient Retrieval of Similar Time Sequences Under Time Warping}.
\newblock \bibinfo{journal}{\emph{ICDE}} (\bibinfo{date}{10} \bibinfo{year}{1999}).
\newblock


\end{thebibliography}
\end{document}